\documentclass[twocolumn, switch]{article} 

\usepackage{preprint}

\usepackage{amsmath, amsthm, amssymb, amsfonts}
\usepackage{algpseudocode}
\usepackage{algorithm, amsmath}
\usepackage[numbers,square]{natbib}
\usepackage[utf8]{inputenc}	
\usepackage[T1]{fontenc}	
\usepackage{xcolor}		
\usepackage[colorlinks = true,
            linkcolor = purple,
            urlcolor  = blue,
            citecolor = cyan,
            anchorcolor = black]{hyperref}	
\usepackage{booktabs} 		
\usepackage{nicefrac}		
\usepackage{microtype}		
\usepackage{lineno}		
\usepackage{float}			

\usepackage{lipsum}		

\usepackage{newfloat}
\DeclareFloatingEnvironment[name={Supplementary Figure}]{suppfigure}
\usepackage{sidecap}
\sidecaptionvpos{figure}{c}

\usepackage{titlesec}
\titlespacing\section{0pt}{12pt plus 3pt minus 3pt}{1pt plus 1pt minus 1pt}
\titlespacing\subsection{0pt}{10pt plus 3pt minus 3pt}{1pt plus 1pt minus 1pt}
\titlespacing\subsubsection{0pt}{8pt plus 3pt minus 3pt}{1pt plus 1pt minus 1pt}

\title{Radiologist-Level COVID-19 Detection Using CT Scans with Detail-Oriented Capsule Networks}

\usepackage{xwatermark}
\newwatermark[firstpage,color=black!90,angle=0,scale=0.28, xpos=0in,ypos=-5in]{*correspondence: \texttt{amobiny@uh.edu}}

\usepackage{authblk}

\author[1,*,$\dagger$]{Aryan Mobiny}
\author[1,2,$\dagger$]{Pietro~A.~Cicalese}
\author[1]{Samira~Zare}
\author[1]{Pengyu~Yuan}
\author[1]{Mohammad S. Abavisan}
\author[3]{Carol C. Wu}
\author[3]{Jitesh Ahuja}
\author[3]{Patricia M. de Groot}
\author[1]{Hien~V.~Nguyen}

\affil[1]{Department of Electrical and Computer Engineering, University of Houston, Houston,
TX, 77004 USA}
\affil[2]{Department of Biomedical Engineering, University of Houston, Houston,
TX, 77004 USA}
\affil[3]{Department of Thoracic Imaging, University of Texas MD Anderson Cancer Center, Houston, TX, 77030 USA}
\affil[$\dagger$]{These authors contributed equally to the completion of this work.}
\affil[*]{corresponding author: \texttt{amobiny@uh.edu}}

\begin{document}

\twocolumn[ 
  \begin{@twocolumnfalse} 
  
\maketitle

\begin{abstract}
Radiographic images offer an alternative method for the rapid screening and monitoring of Coronavirus Disease 2019 (COVID-19) patients. This approach is limited by the shortage of radiology experts who can provide a timely interpretation of these images. Motivated by this challenge, our paper proposes a novel learning architecture, called Detail-Oriented Capsule Networks (DECAPS), for the automatic diagnosis of COVID-19 from Computed Tomography (CT) scans. Our network combines the strength of Capsule Networks with several architecture improvements meant to boost classification accuracies. First, DECAPS uses an Inverted Dynamic Routing mechanism which increases model stability by preventing the passage of information from non-descriptive regions. Second, DECAPS employs a Peekaboo training procedure which uses a two-stage patch crop and drop strategy to encourage the network to generate activation maps for every target concept. The network then uses the activation maps to focus on regions of interest and combines both coarse and fine-grained representations of the data. Finally, we use a data augmentation method based on conditional generative adversarial networks to deal with the issue of data scarcity. Our model achieves 84.3\% precision, 91.5\% recall, and 96.1\% area under the ROC curve, significantly outperforming state-of-the-art methods. We compare the performance of the DECAPS model with three experienced, well-trained thoracic radiologists and show that the architecture significantly outperforms them. While further studies on larger datasets are required to confirm this finding, our results imply that architectures like DECAPS can be used to assist radiologists in the CT scan mediated diagnosis of COVID-19.
\end{abstract}


\vspace{0.35cm}

  \end{@twocolumnfalse} 
] 



\section{Introduction}
\label{sec:introduction}
The Coronavirus Disease 2019 (COVID-19) pandemic has taken more than 100,000 human lives, has cost the world economy several trillion dollars, and has fundamentally changed every aspect of our lives. Early, accurate, and scalable technologies for the screening of patients and monitoring of their treatment progress are crucial for controlling this crisis. At the time of writing this paper, real-time Reverse Transcriptase Polymerase Chain Reaction (rRT-PCR) is the most widely used approach for patient screening \cite{RT-PCR}. Patient specimens are usually collected from their nose or throat, and are then sent to laboratories to generate a diagnosis. The process is time-consuming, complicated, and limited by the supply of test kits, swabs, and test reagents, which has led to logistic challenges due to the large influx of patients.


Computed tomography (CT) of the thorax is a tool frequently used to assess patients with fever and/or respiratory symptoms, and it has been used during this pandemic to triage suspected COVID-19 patients for further testing or hospital admission and to detect treatment response or complications \cite{ai2020correlation,fang2020sensitivity}. Recent studies show that CT scans of COVID-19 patients have typical features that can be recognized by doctors and artificial intelligence models \cite{chung2020ct,ai2020correlation,song2020emerging}. While the role of radiographic images is still under discussion as the crisis unfolds \cite{li2020coronavirus}, they have several important advantages including fast turnaround time (e.g., 15 to 30 minutes for a CT scan) and wide availability at most hospitals. These advantages potentially make radiographic imaging techniques a great complementary tool to rRT-PCR testing with which to fight the COVID-19 pandemic \cite{chua2020role,rubin2020role}.

An important challenge in using radiographic imaging is the shortage of well-trained radiologists who can provide a timely and accurate interpretation of patients' images \cite{mossa2020radiology}. A new disease like COVID-19 with variable imaging manifestations requires radiologists to spend significant amounts of time to constantly update their knowledge and learn new interpretation skills. This becomes even more challenging in remote locations with insufficient access to training resources and poorly-equipped medical facilities. The inability to perform image interpretation for the diagnosis of the disease and assessment of disease severity make these remote regions of the country highly vulnerable when the disease strikes. In addition, the pandemic creates a large influx of patients which sharply increases the radiologists' workload. Addressing the shortage of radiology experts is critical to the diagnostic process and is needed to maximize the potential of radiographic imaging in the diagnosis of COVID-19. 

Motivated by the urgent need, recent work has developed a number of artificial intelligence models for automatic diagnosis or assessment of COVID-19 from CT scans and chest radiographic data \cite{wang2020covidnet,li2020artificial,chaganti2020quantification,xu2020deep}. Unfortunately, training these models requires a large number of annotated examples. This is a difficult requirement given that data are not abundantly available nor properly annotated during the initial period of the pandemic, partly because physicians are too busy to provide clinical feedback. In addition, most prior publications do not make their data publicly available due to privacy and proprietary issues \cite{gozes2020rapid,ai2020correlation,li2020artificial}. Therefore, being able to build robust and accurate artificial intelligence models using a small number of imperfect samples is highly desirable.

Existing work is limited to evaluating artificial intelligence models using classification accuracy on an independent test set. While the results are encouraging, it remains unclear how these models fare against well-trained human radiologists and how they derive their conclusions. Past work with deep learning has demonstrated that deep networks are competitive against human experts on a number of classification tasks for chest radiographs \cite{rajpurkar2017chexnet,irvin2019chexpert}, CT scans \cite{grewal2018radnet}, and magnetic-resonance images \cite{fairbank2017automation}. However, the nature of these studies is significantly different from the one in this paper. In particular, while these studies use large-scale and carefully curated datasets, our study focuses on building artificial intelligence models from a modest and possibly noisy set of training examples. In addition, the pathological conditions in prior studies are well-established while COVID-19 is still a new disease whose characteristics have not been fully understood. Comparing the performance of artificial intelligence algorithms against human experts at the developing stage of a new disease is unique. It can potentially shed light on how well computer and human experts stand up to a new diagnostic challenge. Our results are expected to provide a more complete view of artificial intelligence algorithms, including their strengths and weaknesses in responding to the COVID-19 pandemic. Our paper makes the following contributions:
\begin{itemize}
\item We propose a novel detail-oriented capsule network architecture capable of identifying fine-grained and discriminative image features to classify COVID-19 patients. 
\item Our model employs a data augmentation method based on conditional generative adversarial nets that is meant to deal with the issue of data scarcity.
\item We provide a rigorous evaluation of our models with visualization and clinical input from clinicians. Our extensive experiments on the COVID-19 CT dataset indicate that the proposed models are capable of learning meaningful and discriminative features. 
\item Our paper compares the performances of artificial intelligence models to three well-trained and experienced radiologists. To the best of our knowledge, this is the first attempt to benchmark deep learning models against human experts for COVID-19 diagnosis.
\end{itemize}

The rest of this paper is organized as follows: works related to automatic diagnosis of COVID-19, capsule networks, and data augmentation with generative models are presented in Section \ref{related_work}. Section~\ref{methods} explains the proposed network architecture and the methods applied to address data scarcity. Section~\ref{data} describes the dataset used in this study. Experimental results are presented in Section~\ref{results} and are discussed in Section~\ref{discussion}. Section~\ref{conclusion} concludes the paper with future research directions. All relevant source code and radiologist diagnostic data will be made publicly available.

\vspace{-2mm}

\section{Related work}
\label{related_work}



\noindent
\textbf{Automatic Diagnosis of COVID-19:}  COVID-Net was recently introduced in \cite{wang2020covidnet} for the automatic interpretation of chest radiographs of COVID patients. The network consists of convolutional layers with skip connections similar to residual networks \cite{he2016deep}. Authors showed promising accuracy in classifying X-ray images aggregated from various online sources. \cite{cohen2020covid}. Chaganti \emph{et al.}  \cite{chaganti2020quantification} proposed a system that finds suspicious lung regions in a CT image using a deep reinforcement learning algorithm and then measured the abnormality and severity caused by the virus. Li \emph{et al.}  \cite{li2020artificial} modified a residual network with 50 layers to detect COVID-19 using CT scans. The model was trained on a dataset comprised of 4,356 CT scans collected from multiple hospitals in China. The dataset, however, has not been made available to the public. \cite{xu2020deep} proposed an architecture based on ResNet18 to extract features from CT scans, segment out pathological lung regions which then serve as the input to a classifier for predict COVID-19 disease. The network achieves an overall accuracy of 86.7\%. While these studies demonstrated the great potential of artificial intelligence algorithms in automating COVID-19 diagnosis, most of them do not address the problem of data scarcity in model training, which is a practical constraint at the beginning of a pandemic. More importantly, their evaluations are limited to standard classification metrics on independent test sets. None of the work has compared the performances of artificial intelligence models to radiology experts. Such a comparison coupled with the qualitative visualization of our results will improve our understanding of the strengths and weaknesses of our algorithms while ensuring that high accuracies are not the result of over-fitting.

\vspace{1mm}
\noindent
\textbf{Capsule Networks: } Capsule Networks (CapsNets) \cite{sabour2017dynamic,hinton2018matrix} have demonstrated to be a promising alternative to convolutional neural networks (CNNs). They are capable of encoding part-whole relationships and the relative pose of objects with respect to the surrounding scene \cite{sabour2017dynamic}. CapsNets employ an iterative routing-by-agreement mechanism to determine where to send information which allows capsules to learn viewpoint invariant representations of the data. CapsNets achieved state-of-the-art performance in many tasks including hand-written digit recognition \cite{sabour2017dynamic}, human action detection in videos \cite{duarte2018videocapsulenet}, and cancer screening with volumetric radiology data \cite{mobiny2018fast}. However, CapsNets with dynamic or EM routing experience unstable training when the number of layers increases \cite{mobiny2019automated, rawlinson2018sparse}. It is also unclear how CapsNets select the informative regions within the image to make their decision. Our paper introduces a stable detail-oriented CapsNet architecture that intelligently filters out irrelevant information, effectively capturing both coarse and fine-grained representations within the data.

\vspace{1mm}
\noindent
\textbf{Generative Models for Data Augmentation: }
Acquiring a large amount of training samples that effectively characterize population diversity is necessary for the success of deep neural networks in medical applications. However, COVID-19 imaging data availability, especially during a crisis like the COVID-19 pandemic, is quite limited; high annotation costs, overwhelmed physicians, and time restrictions all contribute to data scarcity. Therefore, generating descriptive and realistic synthetic training samples serves as an enticing solution. Jin \emph{et al.} \cite{jin2018ct} employed a 3D generative adversarial network to artificially generate fake lung CT scans with lung nodules of various dimensions at multiple locations and achieved promising results. Han \emph{et al.}  \cite{han2019synthesizing} proposed a 3D multi-conditional GAN (3D MCGAN) which uses context and nodule discriminators with distinct loss functions to generate realistic and diverse nodules in CT images. Ghorbani \emph{et al.} \cite{ghorbani2019dermgan} proposed DermGAN (implementing the pix2pix architecture), which can translate skin conditions to a synthetic sample with a new size, location and skin color. We propose the use of the pix2pix architecture as a means of augmenting our relatively scarce COVID-19 2D CT dataset due to the quality, high sample stochasticity, and feature fidelity it achieves.

 \begin{figure*}[!t]
\centering
\includegraphics[width=0.9\textwidth]{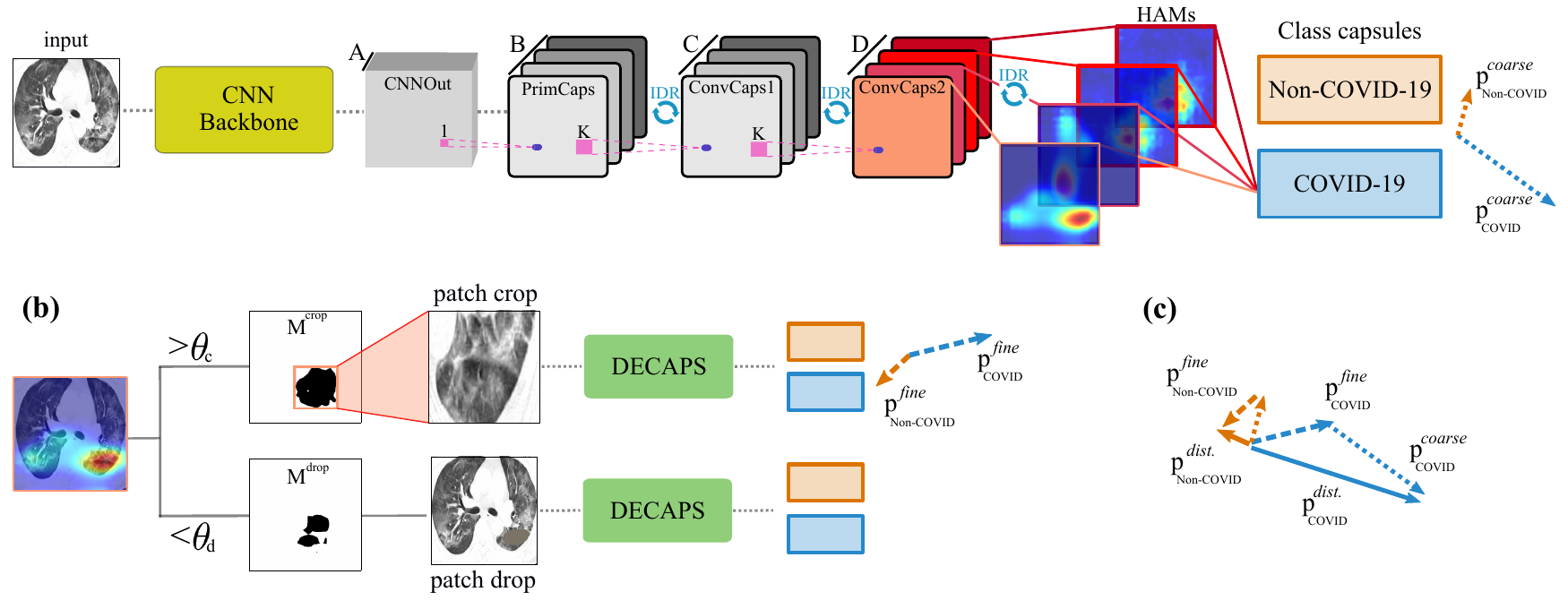}
\caption{\textbf{(a):} Illustration of the DECAPS architecture. Sample head activation maps (HAMs) are presented for the COVID-19 class. \textbf{(b):} Peekaboo training. HAMs are randomly selected to perform patch crop/drop while training. At test time, the average HAM across all heads is used to perform patch crop (with no patch drop required at test time) to obtain the fine-grained prediction. \textbf{(c):} The distillation process to fine-tune the coarse-grained prediction ($\text{p}^{\text{coarse}}$) using the fine-grained prediction ($\text{p}^{\text{fine}}$) to obtain the final distilled predictions ($\text{p}^{\text{dist.}}$).}
\label{model}
\end{figure*}

\section{Methodology}
\label{methods}

The accurate assessment of a radiology image depends on the physician's ability to focus on relevant information while simultaneously considering the context of neighboring regions and the entire scan. It would therefore be preferable to simulate this behavior by using a classification architecture that is capable of deriving conclusions about regions of interest with respect to the greater context of the image. To accomplish this, we propose the Detail Oriented Capsule Network (DECAPS) architecture, which uses two underlying concepts to replicate how a radiologist reviews each scan. First, we use a Capsule Network (CapsNet) classifier, which allows the architecture to intelligently combine information within and around a region of interest to generate a holistic interpretation of each scan. Second, we introduce the ``peekaboo'' training procedure which randomly retrieves or removes areas of interest in the image. When we pass the retrieved ROI and the entire scan through the classifier, the \textit{local} classification power of the architecture is effectively increased by simulating the ability of the radiologist to consider the ROI with respect to the entire scan. Alternatively, we pass the whole scan with the ROI removed; this forces the classifier to recognize multiple ROIs when possible, thus enhancing its \textit{global} classification power.

Radiology data is often comparatively scarce in the context of classification tasks; this is especially true for images derived from novel disease cases like COVID-19. This presents a unique challenge for researchers in how to best utilize their available data to generate robust and reliable models that are able to generalize to the patient population. While popular online augmentation techniques such as image flipping, rotation, and cropping help improve generalization, they fail to minimize the impact of otherwise irrelevant information that is coincidentally correlated to the classification labels. Unique image characteristics such as organ shapes, scan irregularities, and other morphologically irrelevant information can easily distract a classifier trained on a small sample of data, thus harming its ability to generalize. Considering the urgent need for the dependable and high-throughput diagnosis of COVID-19, we believe that leveraging intelligent data augmentation techniques that learn to identify and combine class specific features is fundamental to building an inexpensive but robust COVID-19 classifier. To accomplish this, we propose the use of conditional GANs as a means to model the distribution of the CT image data for each class; this allows us to then generate unique samples from the distribution that can then be passed to our desired classification architecture to improve its generalization ability.

\noindent \textbf{Notations:} Throughout the paper, $r$, $\text{r}$, $\text{R}$, $\mathbf{R}$ represent a scalar, a vector, a 2D matrix, and a tensor (i.e. a higher dimensional matrix; usually a 3D matrix of capsule activations), respectively. Multiplying a transformation matrix and a tensor of poses is equivalent to applying the transformation to each pose.

\subsection{Background on Capsule Networks}

A CapsNet is composed of a sequence of capsule layers, each of which contains multiple discriminative elements called capsules. We can define each capsule as being a group of neurons that effectively encode a pose vector or matrix \cite{sabour2017dynamic, hinton2018matrix}. Suppose that we have a set of capsules in layer $L$ which we denote as $\Omega_L$. We then say that each capsule $i \in \Omega_L$ outputs a pose vector $\text{p}^L_i$; each element in the matrix characterizes the instantiation parameters (such as orientation, size, etc.). The activation probability of a capsule $a^L_i$ indicates the presence of an entity and is implicitly encoded in the capsule as the Frobenius norm of the pose vector. We then generate a vote vector $\text{v}^L_{ij}$ by processing the information from the $i$-th capsule in $\Omega_L$ to the $j$-th capsule in $\Omega_{L+1}$ by using a linear transformation $\text{v}^L_{ij} = \text{W}^L_{ij} \text{p}^L_i$. The pose of a given capsule $j \in \Omega_{L+1}$ is generated using a convex combination of all the votes from child capsules: $\text{p}^{(L+1)}_j=\sum_i r_{ij} \text{v}^L_{ij} $, where $r_{ij}$ are routing coefficients and $\sum_{i} r_{ij}=1$. We generate these coefficients through a dynamic routing algorithm which increases the routing coefficient $r_{ij}$ when the corresponding vote vector $\text{v}_{ij}^L$ and $\text{p}_j^{L+1}$ are similar to each other \cite{sabour2017dynamic}. This process weights the output of each child capsule to be passed to the corresponding parent capsules. The network therefore constructs a transformation matrix for all capsule pairs to encode the part-whole relationships (i.e. parsing a semantic tree between capsules) while retaining the geometric information of the input data. This property has prompted several research groups to develop new capsule designs and routing algorithms \cite{mobiny2018fast, ahmed2019star, mobiny2019automated,kosiorek2019stacked}.

\subsection{Detail Oriented Capsule Networks (DECAPS)}

As mentioned in the previous section, the pose of parent capsules within the vanilla CapsNet architecture are derived from the votes of all corresponding children capsules. This can have a negative effect of the performance of the classifier; while some child capsules pass useful information from regions of interest, others pass noise derived from non-descriptive areas. To address this limitation, we implemented a unique CapsNet architecture, loss function, and inverted routing mechanism which increases the weight of votes derived from ROIs. This effectively improves the quality of the input information being passed to each parent, thus strengthening the networks ability to build part-whole relationships. Note that given the relatively small sample size of our dataset, we chose to utilize pose matrices to decrease the number of training parameters to prevent the early over-fitting that is associated with a large number of parameters \cite{hinton2018matrix}. We also took inspiration from the Transformers architecture described by Vaswani \emph{et al.} and grouped capsules into grids called \textit{Capsule Heads}, as shown in Fig. \ref{model} (a). Each capsule head is designed to route information to parent capsules independently of the other capsule heads; we encourage this by forcing each head to share a transformation matrix $\text{W}^L_{ij}$ between all capsules for each output class. This differs from the vanilla CapsNet architecture which uses one transformation matrix per capsule per class, and effectively reduces the number of trainable parameters by an order of head size (or the number of capsules within a head). This allows us to both train our network with large images and increase the number of capsule layers used (thus increasing abstraction and the complexity of the features captured by the classifier).

We let $\mathbf{P}^L_i \in \mathbb{R}^{h_L \times w_L \times d_L}$ represent the pose matrix of the capsules of the $i^{th}$ head where $h_L$ and $w_L$ represent the height and width of the head respectively, while $d_L$ denotes the capsule dimension (i.e. the number of hidden units grouped together to yield the capsules in layer $L$). We change $i$ to represent the index of the head in our architecture as opposed to its previous definition in our vanilla CapsNet description (the child capsule index). We take the Frobenius norm of each capsule pose matrix to represent the existence probability of the desired semantic category; these values are squashed through a non-linear squash function to ensure that they fall between zero (not present) and one (present) \cite{sabour2017dynamic}. We then define the votes from the capsules of the $i^{th}$ head to the $j^{th}$ parent capsule as $\mathbf{V}^L_{ij}=\text{W}^L_{ij}\mathbf{P}^L_i$. For each coordinate within a capsule head, we take the relative coordinates of each capsule (row and column) and add them to the final two entries of the vote vector, allowing us to preserve the capsules location \cite{hinton2018matrix}. At this point, the generated votes are processed by the inverted routing mechanism to effectively parse each semantic tree present in the data.

\begin{algorithm*}[t]
\caption{Inverted Dynamic Routing (IDR). $i$ and $j$ are the indices of capsule heads in layer $L$ and $L+1$ respectively.}
\label{algorithm}
\begin{algorithmic}[1]
\Procedure{IDR}{$\mathbf{V}^{L}_{ij}, n^L_{\text{iter}}$}
\Comment{given the votes and number of routing iterations}
\State $\text{R}_{ij}^{\text{pre}, L}\gets 0$ \Comment{initialize routing coefficients}
\For{$n^L_{\text{iter}}$ iterations}
\State $\text{R}^L_{ij}\gets \text{{\fontfamily{lmtt}\selectfont softmax}} (\text{R}_{ij}^{\text{pre}, L})$ \Comment{Softmax among capsules in head $i$}


\State $\mathbf{\Tilde{A}}^L_{ij} \gets \text{R}^L_{ij} \odot \mathbf{V}^{L}_{ij}$
\Comment{$\odot$ is the Hadamard product}

\State $\text{P}_j^{L+1} \gets \text{{\fontfamily{lmtt}\selectfont squash}}(\sum_i \sum_{xy} \mathbf{\Tilde{A}}^L_{ij})$
\Comment{$\sum_{i}$ and $\sum_{xy}$ are sums over heads and locations, respectively}

\State $\text{R}_{ij}^{\text{pre}, L} \gets \text{R}_{ij}^{\text{pre}, L} + \text{P}_j^{(L+1)} .  \mathbf{V}^{L}_{ij}$
\EndFor

\State $\text{A}^L_{ij} \gets \text{{\fontfamily{lmtt}\selectfont length}} (\mathbf{\Tilde{A}}^L_{ij})$
\Comment{$\text{{\fontfamily{lmtt}\selectfont length}}$ computed by Eq. \eqref{eq1}}
\State \textbf{return} $\text{P}_j^{L+1}, \text{A}^L_{ij}$
\EndProcedure
\end{algorithmic}
\end{algorithm*}

\begin{figure}[!b]
\centering
\includegraphics[width=0.45\textwidth]{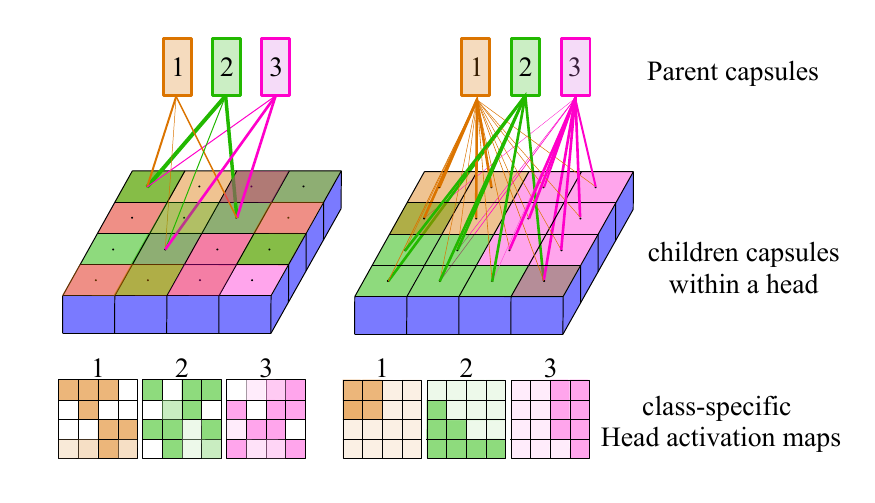}
\label{idr}
\vspace{-5mm}
\caption{Dynamic (left) vs. Inverted dynamic routing (right). Capsule heads are shown in different colors.}
\end{figure}

\vspace{1mm}
\noindent
\textbf{Inverted Dynamic Routing: }
Dynamic routing \cite{sabour2017dynamic} is a \textit{bottom-up} approach which forces higher-level capsules to compete with each other to collect lower-level capsule votes. We propose an inverted dynamic routing (IDR) technique which implements a \textit{top-down} approach, effectively forcing lower-level capsules to compete for the attention of higher-level capsules (see Fig. \ref{idr}). During each iteration of the routing procedure, we use a softmax function to force the routing coefficients between all capsules of a single head and a single parent capsule to sum to one (see Algorithm \ref{algorithm}). The pose matrix of the $j^{th}$ parent capsule, $\text{P}^{L+1}_j$, is then set to the squashed weighted-sum over all votes from the earlier layer (line 6 in Algorithm \ref{algorithm}). Given the vote map computed as $\mathbf{V}^L_{ij}=\text{W}^L_{ij}\mathbf{P}^L_i \in \mathbb{R}^{h_L\times w_L\times d_{L+1}}$, the proposed algorithm generates a routing map $\text{R}^L_{ij} \in \mathbb{R}^{h_L\times w_L}$ from each capsule head to each output class. The voting map describes the children capsules' votes for the parent capsule's pose. The routing map depicts the weights of the children capsules according to their agreements with parent capsules, with winners having the highest $r^L_{ij}$. We combine these maps to generate head activation maps (or HAMs) following

\vspace{-2mm}
\begin{equation}
\label{eq1}
\text{A}^L_{ij} = \Big({\sum\nolimits_{d} (\mathbf{\Tilde{A}}^L_{ij})^2}\Big)^{1/2}, \quad \text{where} \quad \mathbf{\Tilde{A}}^L_{ij} = \text{R}^L_{ij} \odot \mathbf{V}^{L}_{ij}
\end{equation}

\noindent
in which $\text{A}^L_{ij}$ is the head activation map from the $i^{th}$ head in layer $L$ to the $j^{th}$ head in layer $L+1$, and $\sum_d$ is the sum over $d_{L+1}$ channels along the third dimension of $\mathbf{V}_{ij}^L$. In the output of the last convolutional capsule layer, $\text{A}^L_{ij}$ highlights the informative regions within an input image corresponding to the $j^{th}$ class, captured by the $i^{th}$ head. IDR returns as many activation maps as the number of capsule heads per output class (see Fig. \ref{idr}). Class-specific activation maps are the natural output of the proposed framework, unlike CNNs which require the use of additional modules, such as channel grouping, to cluster spatially-correlated patterns \cite{zheng2017learning}. We utilize the activation maps to generate ROIs when an object is detected. This effectively yields a model capable of weakly-supervised localization which is trained end-to-end; we train on the images with categorical annotations and predict both the category and the \emph{location} (i.e. mask or bounding box) for each test image. This framework is thus able to simultaneously generate multiple ROIs within the same image that are associated with different medical conditions.

\vspace{1mm}
\noindent
\textbf{The DECAPS Architecture:}
The overall architecture of the proposed DECAPS is shown in Fig. \ref{model}. It uses a ResNet \cite{he2016deep} with three residual blocks as the base network which outputs 1024 feature maps, followed by a $1 \times 1$ convolutional layer with $A=512$ channels and a ReLU nonlinearity. All the other layers are capsule layers starting with the primary capsule layer.  The $4 \times 4$ pose matrix of each of the B primary
capsule heads is a learned linear transformation of the output of all the lower-layer ReLUs centered at that location. The primary capsules are followed by two convolutional capsule layers with C and D capsule heads and kernels of size $K=3$ and stride $s=1$. We selected $B=C=D=32$ capsule heads for the capsule layers, and used the proposed inverted dynamic routing (IDR) mechanism to rout the information between the capsules. The last layer of convolutional capsules is linked to the final dense capsule layer which has one capsule per output class. The Frobenius norm of the pose matrices of the output capsules are used to determine the predicted class.

\vspace{1mm}
\noindent
\textbf{Loss Function: }We use spread loss to enforce the pose matrices of the top-level capsule $j$ to have a large Frobenius norm if and only if the object of the corresponding class exists in the image \cite{hinton2018matrix}. The total spread loss is the sum of the losses for each output capsule as given by

\vspace{-2mm}
\begin{equation}
\label{eq2}
L_{\text{spread}}=\sum\nolimits_{j\neq t} L_{j} = \sum\nolimits_{j \neq t} \text{max}(0, m - (a_t-a_j))^2
\end{equation}

\noindent which directly maximizes the distance between the activation of the target class, $a_t$, and the activation of the other classes, $a_i$, using the margin $m$. In other words, it penalizes the model if the distance between the $a_t$ and $a_i$ is less than the margin $m$. We initialize the margin to $0.2$ and linearly increase it by adding $0.1$ to its value every other epoch, which reduces the incidence of dead capsules in the hidden layers.

\vspace{1mm}
\noindent
\textbf{Activation-Guided Training (Peekaboo): }
To further promote DECAPS to focus on fine-grained details, we propose the Peekaboo strategy for capsule networks. Our strategy boosts the performance of DECAPS by forcing the network to look at all relevant parts for a given category, not just the most discriminative parts. It is inspired by the Hide and Seek \cite{singh2017hide} mechanism in which image patches are randomly dropped to encourage the network to look for other relevant parts. Peekaboo, however, uses the HAMs to guide the network's attention process. For each training image, we randomly select an activation map $\text{A}_{ij}$ for each recognized category. Each map is then normalized in the range $[0, 1]$ to get the normalized HAM, $\text{A}^{*}_{ij} \in \mathbb{R}^{h_L\times w_L}$. We then enter a two step process: patch cropping, which extracts a fine-grained representation of the ROI to learn how to encode details, and patch dropping, which encourages the network to attend to multiple ROIs. In patch cropping, a mask $\text{M}^{\text{crop}}_{ij} \in \mathbb{R}^{h_L\times w_L}$ is obtained by setting all elements of $\text{A}^{*}_{ij}$ which are less than a cropping threshold $\theta_c \in [0, 1]$ to 0, and 1 otherwise. We then find the smallest bounding box which covers the entire ROI, and crop it from the raw image (Fig. \ref{model} (b)). It is then upsampled and fed into the network to generate a detailed description of the ROI. During the patch dropping procedure, $\text{M}^{\text{drop}}_{ij}$ is used to remove the ROI from the raw image by using a dropping threshold $\theta_d \in [0, 1]$. The new patch-dropped image is then fed to the network for prediction. This encourages the network to train capsule heads to attend to multiple discriminative semantic patterns. Note that we flatten the output prediction matrices to form vectors (as shown in Fig. \ref{model}) whose magnitude represents the existence probability of the corresponding class. 

At test time, we first input the whole image to obtain the coarse prediction matrices.  ($\text{p}^{\text{coarse}}_j$ for the $j^{th}$ class) and the HAMs $\text{A}_{ij}$ from all capsule heads. We then average all maps across the heads, crop and upsample the ROIs, and feed the regions to the network to obtain the fine-grained prediction vectors ($\text{p}^{\text{fine}}_j$). The final prediction $\text{p}^{\text{dist.}}_j$, referred to as distillation (Fig. \ref{model} (c)), is the average of the $\text{p}_j^{\text{coarse}}$ and $\text{p}_j^{\text{fine}}$. The Peekaboo strategy can also be implicitly interpreted as an online data augmentation technique; it intelligently alters the input images (by either occluding or passing up-sampled crops of regions of interest) to synthetically increase the number of samples passed through the network. We can therefore say that this feature helps to mitigate the effects of data scarcity.

\begin{figure*}[!t]
\centering
\includegraphics[width=0.8\textwidth]{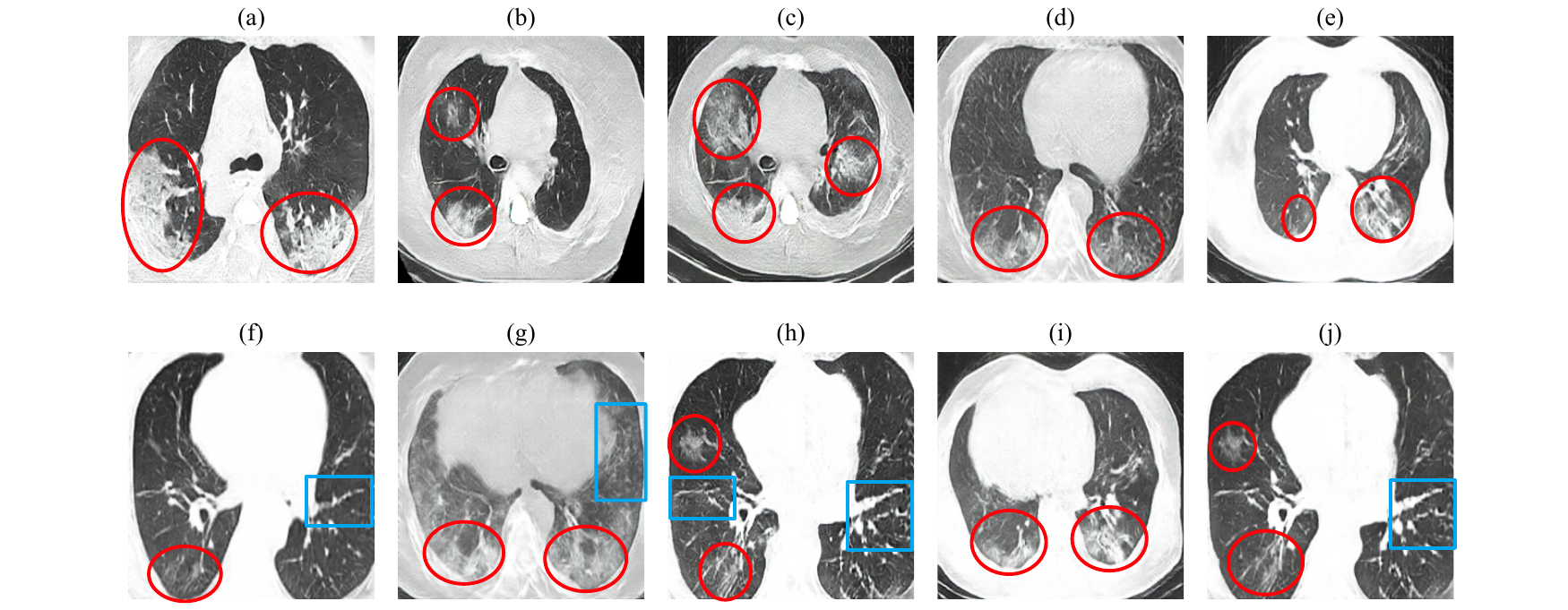}
\caption{Qualitative assessment of COVID-19 CT pix2pix deep fakes. Red circles highlight characteristics consistent with the diagnosis of COVID-19, while blue squares correspond to features of non-COVID-19 diseases. We hypothesize that cases containing both kinds of features (``artificial co-morbidity'' cases, (f), (g), (h), and (j)) help to enforce the unique characteristics of COVID-19.}
\label{fig_sim}
\end{figure*}

\subsection{Image-to-Image Conditional GAN Data Augmentation}

We propose the use of the Image-to-Image (pix2pix) conditional GAN architecture as a means to augment our otherwise scarce COVID-19 CT scan dataset \cite{isola2017image}. This architecture learns the mapping of both the input set (non-COVID-19 CT scans) and the output set (COVID-19 CT scans) by using a \textit{learned} loss function, thus not requiring complex loss formulations. This allows us to learn the data distribution for each class; characteristics that make each class unique are effectively encoded and then sampled to create unique cases. We selected the pix2pix architecture due to its ability to retain stochasticity in the generated data, a feature that is lost in other conditional GAN architectures; this randomness is critical to the generation of unique cases that enrich our data. To accomplish this, we drew from the original implementation of pix2pix, substituting the Gaussian noise typically introduced in conditional GAN architectures with train-test dropout. Our generative model was also designed using the characteristic ``U-net'' architecture, where each layer $i$ has a skip connection to layer $n-i$, with $n$ being equal to the total number of layers \cite{ronneberger2015u}. This effectively prevents the network from minimizing the low-level information being passed, thus improving the overall quality of the output. Finally, we mitigate the negative effect the L1 loss has on image ``crispness'' by restricting the discriminative model's attention to local image patches (i.e. the PatchGAN architecture). This effectively allows us to model high-level information while relying on the L1 loss to force low-level correctness, thus modeling the images as a Markov random field (i.e. simulating style loss). Additionally, we note that the PatchGAN architecture significantly reduces the number of trainable parameters and can therefore be applied to larger images (such as full resolution CT scans).

\section{COVID-19 CT dataset}
\label{data}

In this paper, an open-source dataset prepared by Zhao \emph{et al.}  \cite{zhao2020COVID-CT-Dataset} is utilized to train and test the proposed model. It contains a total of 746 chest CT images, which are divided into two classes, namely COVID-19 and non-COVID-19. A pre-processed version of the dataset is available at \url{https://github.com/UCSD-AI4H/COVID-CT}.

The dataset was created by collecting images from papers related to COVID-19 published in medRxiv, bioRxiv, NEJM, JAMA, Lancet, and other impactful journals. Then, the images were classified according to the figure captions describing the clinical findings in the papers. All copyrights of the data belong to the authors and publishers of these papers. 349 CT images were labeled as COVID-19 and 397 CT images as non-COVID-19. The heights of these images vary; they range between 153 and 1853 pixels (average of 491 pixels), while their widths range between 124 and 1458 pixels (average of 383 pixels).  We use nearly 85\% of data for training/validation purposes and the rest for model testing. Specifically, the dataset is divided into a training set of 625 CT images (286 positives, 339 negatives) and a test set of 105 CT images (47 positives, 58 negatives). 

\begin{table*}[!t]
\centering
\caption{Prediction performance of models trained on the CT dataset. For each model, average ($\pm$ std.) performance measure is reported over the best 5 trained model checkpoints.}
\resizebox{2\columnwidth}{!}{
\begin{tabular}{lccccccc}
\hline
  & \textbf{params. (M)} & \textbf{Precision} & \textbf{Recall} & \textbf{Specificity} & \textbf{Accuracy} & \textbf{F1-score} & \textbf{AUC}\\ \hline
  
Inception-v3 \cite{szegedy2016rethinking} & \;$25.1$ & \;$\mathbf{0.844(\pm0.039)}$ & \;$0.740(\pm0.119)$  & \;$0.853(\pm0.056)$ & \;$0.819(\pm0.028)$ & $0.781(\pm0.052)$ & $0.894(\pm0.029)$\\
 
DenseNet121 \cite{huang2017densely} & $7.0$ & $0.815(\pm0.063)$  & $0.794(\pm0.098)$ & $0.839(\pm0.074)$ & $0.825(\pm0.027)$  & $0.801(\pm0.038)$ & $0.903(\pm0.022)$\\

ResNet50 \cite{he2016deep}  & $23.5$  & $0.7539(\pm0.036)$ & $0.849(\pm0.121)$  & $0.769(\pm0.068)$ & $0.808(\pm0.039)$ & $0.795(\pm0.057)$ & $0.880(\pm0.030)$\\ \hline

\textbf{DECAPS} & $9.8$ & $0.825(\pm0.038)$ & $0.854(\pm0.079)$  & $\mathbf{0.860(\pm0.074)}$ & $0.832(\pm0.031)$ & $0.837(\pm0.031)$ & $0.927(\pm0.017)$\\

\textbf{DECAPS+Peekaboo}  & $9.8$ & $\mathbf{0.843(\pm0.024)}$  & $\mathbf{0.915(\pm0.057)}$ & $0.852(\pm0.040)$ & $\mathbf{0.876(\pm0.010)}$   & $\mathbf{0.871(\pm0.019)}$ & $\mathbf{0.961(\pm0.009)}$\\ \hline
\end{tabular}
}
\label{table:result_ct}
\end{table*}

\section{Experimental Results}
\label{results}

\noindent
\textbf{Implementation details: }
All COVID-19 CT deep fakes were generated using the pix2pix architecture; images were re-scaled to $286\times286$ and then cropped following the PatchGAN mechanism to $256\times256$. Each input (non-COVID-19) and output (COVID-19) was selected at random during the training procedure with replacement; this was done to maximize the stochasticity of the resulting augmented dataset. All hyperparameter selections were made to ensure that the generated output image was as sharp as possible for interpretation by experienced radiologists. We implemented the U-net 256 architecture as the image generator model using an Adam optimizer ($\beta_1$ = 0.5) and an initial learning rate of $0.0002$ with a linear learning rate policy. All models were trained for $200$ epochs with a batch size of $2$; learning rate was kept constant for the first $100$ epochs and then allowed to decay linearly to zero for the last $100$ epochs. Finally, the L1 to discriminator loss weight ratio was $10$ following the original implementation of pix2pix.

For DECAPS, the best performance was achieved using 3 routing iterations between the capsule layers. The Peekaboo hyperparameters were set as $\theta_c=0.5$, and $\theta_d=0.3$. The network is trained using the Adam optimizer with $\beta_1=0.5$, $\beta_2=0.999$, a learning rate of $10^{-4}$ (fixed for the duration of training) and a fixed batch size of 16. Input images are fed into the network with size $448\times 448$ pixels yielding a $24 \times 24$ capsule map per head for the last convolutional capsule layer.

\vspace{1mm}
\noindent
\textbf{COVID-19 CT Deep Fakes: }
A total of $900$ COVID-19 CT deep fakes were generated using the COVID-19 and non-COVID-19 samples; these images were later added to the original COVID-19 training set during the classification phase. We trained the pix2pix model until the resulting lung tissue information was sharp and qualitatively consistent with the training set images. To determine the semantic quality of the generated samples, we presented a small subset of the generated images to three experienced radiologists for evaluation. In Fig. \ref{fig_sim}, we note that the collaborating radiologists highlighted regions consistent with the manifestation of COVID-19 (red outlines) and areas that appear to resemble other diseases (blue outlines). All of the images in the subset were considered consistent with COVID-19 by at least one of the three radiologists. Given that the images in the non-COVID-19 set were not always control cases, some of the generated cases contained characteristics consistent with both COVID-19 and other diseases.

  \begin{table}[!b]
\centering
\caption{Mean AUC ($\pm$std.) over the top five models trained on the CT dataset to show the effect of each component and their combinations.}
\resizebox{\columnwidth}{!}{
\begin{tabular}{c|c|c|c|c|c}
\hline
 IDR & Pix2Pix & Patch Drop  & Patch Crop & Distillation & AUC \\ \hline
&  &  &  &  & \;$0.844(\pm0.031)$\; \\
\checkmark&  &  &  &  & \;$0.857(\pm0.027)$\; \\
& \checkmark &  &  &  & \;$0.874(\pm0.041)$\; \\
\checkmark& \checkmark &  &  &  & \;$0.927(\pm0.017)$\; \\ \hline
\checkmark& \checkmark & \checkmark &  &   & \;$0.938(\pm0.022)$\; \\
\checkmark& \checkmark &  & \checkmark &   & \;$0.944(\pm0.017)$\; \\
\checkmark& \checkmark & \checkmark & \checkmark &   & \;$0.952(\pm0.013)$\; \\ \hline
\checkmark& \checkmark & \checkmark & \checkmark & \checkmark  & \;$0.961(\pm0.009)$\; \\ \hline
\end{tabular}
}
\label{table:ablation_ct}
\end{table}

\begin{figure}[!b]
\centering
\includegraphics[width=0.5\textwidth]{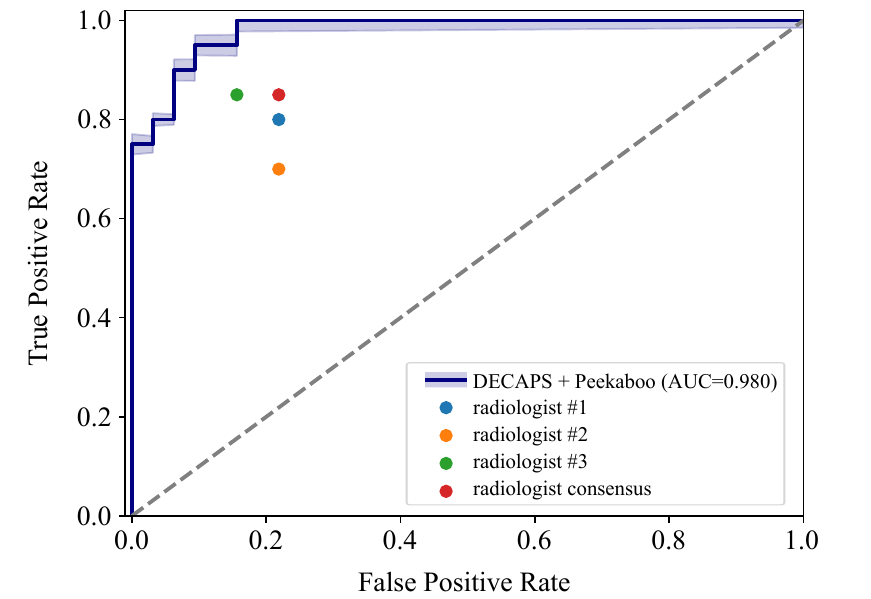}
\caption{Performance comparison between the DECAPS+Peekaboo pix2pix augmented model performance and radiologist performance on the Alpha-Test set. We note that the performance of the radiologists (and their consensus performance) is significantly lower than the architecture.}
\label{roc}
\end{figure}

\vspace{1mm}
\noindent
\textbf{Architecture Comparison and Ablation Study: }
To compare the effectiveness of DECAPS to other traditional classification architectures, we trained the Inception-v3, DenseNet121, and ResNet50 architectures with the pix2pix augmented COVID-19 CT dataset. In Table \ref{table:result_ct}, we note that DECAPS outperforms or matches all other architectures in all six statistical measures. The sharp increase in recall we observe indicates that that the DECAPS architecture was more sensitive to COVID-19 cases when coupled with the Peekaboo training procedure. We then performed an ablation study to understand how inverted dynamic routing, pix2pix data augmentation, patch dropping, patch cropping, distillation, and their combination impacted the overall performance of the CapsNet classifier (as shown in Table \ref{table:ablation_ct}). The sharp increase in performance attributed to the pix2pix image augmentation procedure implies that the technique partially addressed the data scarcity issues we mentioned in section \ref{methods}. We note that each added component of the DECAPS architecture improves results, ultimately yielding a final AUC value of $96.1\%$.

\begin{figure*}[!t]
\centering
\includegraphics[width=0.95\textwidth]{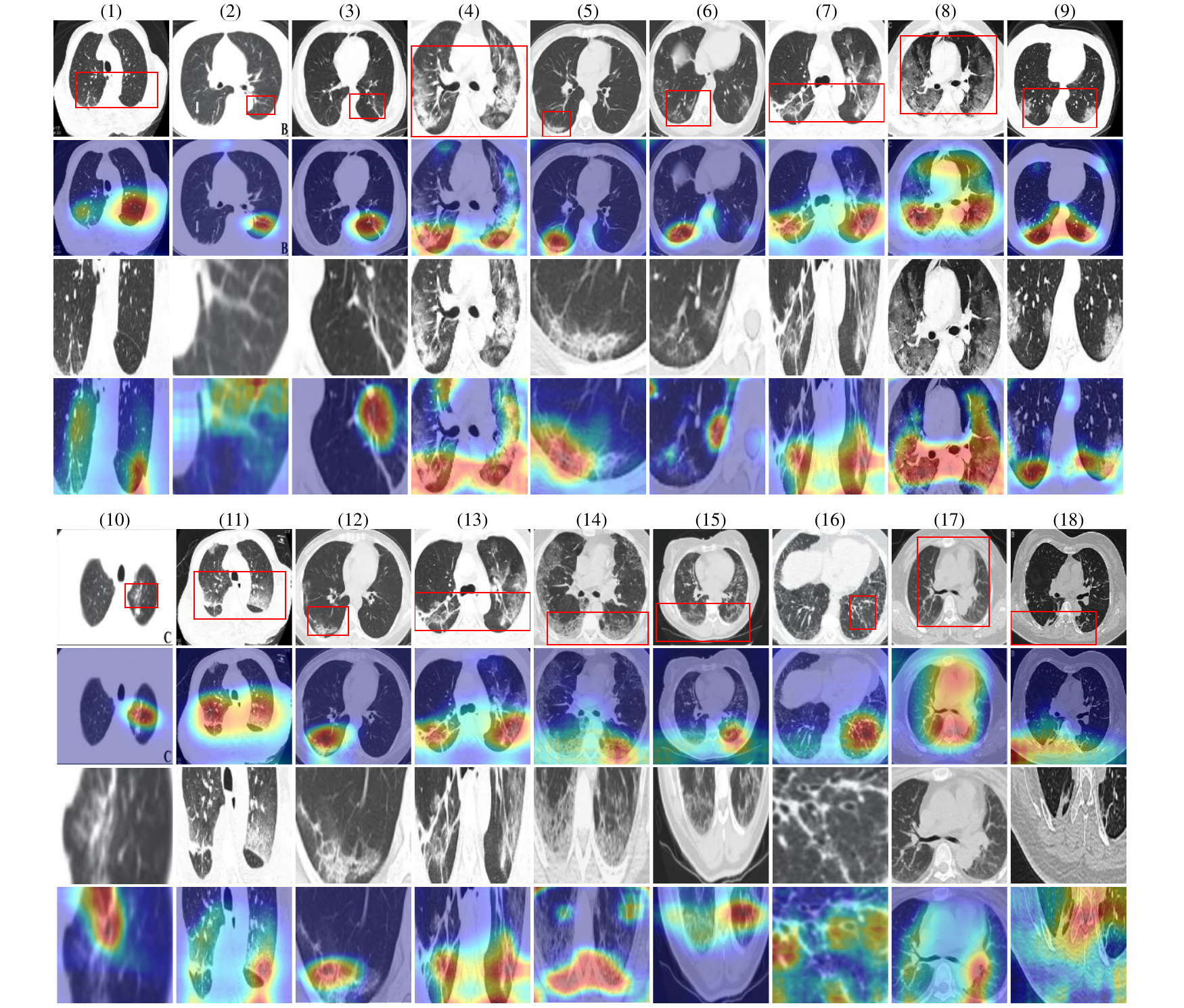}
\caption{Samples describing the qualitative performance of the DECAPS architecture. Each column of 4 images presents the predicted bounding box on the input image, the coarse-grained class activation map, the cropped ROI, and the fine-grained class activation map, in that order. Samples (1)-(13) correspond to true positive cases, samples (14)-(16) correspond to true negative cases, and samples (17)-(18) correspond to failure cases.}
\label{HAMs}
\end{figure*}

\vspace{1mm}
\noindent
\textbf{Comparison to Thoracic Radiologists: }
To evaluate the diagnostic accuracies of the human radiologists and compare them to that of DECAPS, we created a high-quality test set (called Alpha-Test) by manually removing low-resolution images from the original test set and performing visual inspection to ensure that these images do not have artifacts such as radiographic annotations or texts that potentially bias the human radiologists. Our Alpha-Test eventually has 20 COVID positive cases and 32 non-COVID cases. Radiologists independently perform diagnosis on each image without knowing their ground truth information. We generate consensus diagnoses by computing the majority voting (e.g., diagnosis agreed by the majority of radiologists) to simulate the decision from a panel of radiologists. Fig. \ref{roc} shows the Receiving-Operating Characteristic (ROC) curve of our network compared to radiologists' performances. DECAPS + Peekaboo yields the best results with an area under curve (AUC) of $98\%$, and significantly outperforms all 3 radiologists. At a $15.64\%$ false positive rate, the network achieves a $95\%$ true positive rate compared to $85.11\%$ true positive rate of the best radiologist. We also observe that the performance of the radiologist consensus is better than that of 2 radiologists and worse than the best radiologist; at a $21.94\%$ false positive rate, DECAPS + Peekaboo achieves a perfect true positive rate while the consensus achieves an $85.11\%$ true positive rate.

\vspace{1mm}
\noindent
\textbf{DECAPS Visualization and Interpretation:} We generated HAMs for all of the correctly classified COVID-19 cases (Fig. \ref{HAMs}, cases 1-13) to visualize the information being used to generate the COVID-19 label. While most of the HAMs indicate that the correct features are being used, we note that sample 2 is classified using information that is not indicative of COVID-19 (a seemingly normal vascular branch). The samples correctly labeled as non-COVID-19 (Fig. \ref{HAMs}, 14-16) show that the classifier was able to recognize disease chronicity, which is not indicative of acute infection. Finally, we note that both sample 17 and 18 from the misclassified sample group were classified using information that was irrelevant to the classification task.

\section{Discussion}
\label{discussion}


COVID-19 has caused the world significant suffering; the widespread outbreak of the disease has effectively crippled the world healthcare and economic systems, demanding decisive action. While frightening, research shows that the rapid identification and isolation of COVID-19 cases can help to efficiently prevent spread \cite{anderson2020will}. While we would ideally use rRT-PCR to evaluate all potential cases, logistical issues have prevented this technique's widespread adoption and use \cite{cohen2020countries}. It would, therefore, be advantageous for hospitals to make use of resources that they already have; for instance, CT scans are used frequently to triage suspected COVID-19 patients. We were interested in developing a classification architecture that could improve upon the current state of the CT derived diagnostic performance in the context of COVID-19. To accomplish this, we combined a state-of-the-art conditional GAN (pix2pix) data augmentation technique and a novel CapsNet architecture (DECAPS) that attempts to mimic the diagnostic procedures commonly used by radiologists.

Data scarcity is a limiting and difficult problem to solve in the context of AI; while classifiers can be improved and optimized to better understand the information they process, the data itself must adequately reflect some desired concept to be modeled. During a crisis like the COVID-19 pandemic, spending a significant amount of time to collect a well-vetted and large dataset is simply not feasible. This data scarcity is an important roadblock that must be rapidly surmounted to address the automation needs of the medical community in time. We were interested in leveraging the power of conditional GANs to help accomplish this; as has been shown in previous works, ``deep fakes'' could be used to increase the stochasticity of a dataset while retaining some underlying target concept. One can imagine that the generated samples resemble the examples produced by an educator; drawings from a medical textbook (while clearly fake) help to improve the understanding of the learning student. We found that the Image-to-Image (or pix2pix) architecture produced results that were consistent with what we observed in the COVID-19 CT dataset. To ensure that the generated samples captured the underlying concepts that distinguish COVID-19 scans from non-COVID-19 scans, we presented a small sample of the produced cases to three experienced radiologists (as shown in Fig. \ref{fig_sim}). While all three radiologists recognized that the samples were fake, each image was identified as having features consistent with COVID-19 by at least one of the physicians. Interestingly, they noted that several of the samples had features characteristic of other diseases (we refer to these cases as the ``artificial co-morbidity'' cases). For example, Fig. \ref{fig_sim} (g) contains both tree-in-bud opacities in the left lung (blue circle, indicative of non-COVID-19) and ground glass and consolidation (red circles, indicative of COVID-19). Due to the positive effect these samples had on performance, we hypothesize that these cases helped to enforce the unique features of COVID-19 when training the DECAPS classifier.

The success of a classification architecture in radiography imaging tasks is dependent on the ability of the classifier to mimic the behavior of the radiologist. Being conscious of both the entire scan while focusing on a lesion (or lesions) of interest highlights the skill required for the accurate assessment of radiographic data. Most classification architectures are not designed to encourage this kind of behavior; typically, images are simply passed through the classification mechanism without explicitly being forced to consider details with respect to global information. The DECAPS architecture is designed to consider these relationships; it combines fine-grained and coarse information, generating classifications that are derived from the complex relationships between details and the big picture. Based on the generated performance metrics shown in Table \ref{table:result_ct}, we see that the DECAPS architecture succeeds in capturing the characteristics of COVID-19. The sharp increase in recall attributed to the DECAPS architecture implies that the classifier was able to achieve a relatively profound understanding of the features unique to COVID-19. Our ablation study also shows how each unique component in the architecture contributes to this understanding, eventually reaching an AUC value of $96.1\%$ (as shown in Table \ref{table:ablation_ct}). The advantages of this architecture are most evident when compared to the performance of three experienced radiologists; at a false positive rate of $15.64\%$, the top performing radiologist achieves a true positive rate of $85.11\%$ while DECAPS+Peekaboo achieves a true positive rate of $95\%$ (as shown in Fig. \ref{roc}). We also note that the performance of the radiologist consensus is better than that of 2 of the participating radiologists and worse than the best radiologist; at a $21.94\%$ false positive rate, DECAPS+Peekaboo achieves a perfect true positive rate while the consensus achieves an $85.11\%$ true positive rate. Although further experimentation is required, these results imply that DECAPS could potentially match or exceed radiologist performance.

Inspection of all DECAPS HAMs by three experienced radiologists revealed that the classifier made use of information that was relevant to the classification of COVID-19 in most cases (Fig. \ref{HAMs}, (3)-(13)). We note that the classifier was sensitive to peripheral ground-glass opacities which are considered a hallmark of COVID-19 in thoracic CT scans ((5), (6), (9), and (12) as examples). These results indicate that coupling the DECAPS architecture with the augmented dataset effectively addressed most of the underlying data scarcity issues. However, upon observation of sample (2), we note that the architecture used a seemingly normal vascular branch in the left lung to generate its correct COVID-19 classification. We hypothesize that the classifier mistook this feature as a linear band with adjacent ground-glass opacity, which would be indicative of late-stage COVID-19. Similarly, the architecture correctly detected a linear band with adjacent ground-glass opacity in sample (3). It is important to note that the classifier did not ``cheat'' by using the annotation in the right lung of sample (2) that highlighted the peripheral ground-glass opacity, proving that the architecture was seeking COVID-19 specific information. Another important case to consider is sample (1); we note that the classifier focused on a fissure in the left lung (a normal anatomical feature) while simultaneously considering the linear band with adjacent ground-glass opacities from the right lung. As mentioned earlier with sample (2), we believe that the classifier mistook the fissure as a feature of COVID-19. Interestingly, upon further examination, one of the radiologists changed their diagnosis based on the features being highlighted in the right lung; this highlights the importance of incorporating information from multiple regions of the scan to reach the correct diagnosis. This interaction between the physician and the AI system characterizes the advantages of having an AI-physician diagnostic team. 

In samples (14)-(16), we observe cases where the classifier correctly identified traction bronchiectasis as a feature of a chronic fibrotic process, and thus non-COVID-19, despite the presence of ground-glass opacities. Finally, we see two failure cases in samples (17) and (18); while (17) is a non-COVID-19 sample with centrally located ground-glass opacities, sample 18 is a late-stage COVID-19 case mislabeled as non-COVID-19. We hypothesize that the dataset is not fully illustrative of late-stage COVID-19 features due to the nature of the disease and the low supply of descriptive data. The majority of COVID-19 publications have thus far focused on the early presentation of the disease, while patients improving or recovering from the disease usually do not require CT scans, making it difficult to acquire late-stage data.


At the time of writing this paper, healthcare professionals are still fighting against COVID-19 around the world. High-quality data and ground truth annotations are not available in abundance, partly because physicians are too busy taking care of patients to provide clinical feedback. As a result, our study is limited to small training and test sets which potentially fail to represent the true patient population. Further studies on large-scale, multi-institutional, and well-curated datasets are important to confirm our findings. This being said, our results are promising; although our dataset was relatively scarce, we exceeded radiologist performance on a high-quality subset of our testing data. More importantly, our model was used by the radiologists to review and revise their decisions, symbolizing the importance of visualization. Given the severity and sense of urgency instilled by the COVID-19 pandemic, our results suggest that radiologists should utilize COVID-19 CT classifiers like DECAPS to review each case they encounter.

\section{Conclusion}
\label{conclusion}
In this paper, we present Detail Oriented Capsule Networks (DECAPS), a unique capsule network architecture that is meant to mimic the complex behaviors of radiologists in the COVID-19 CT classification task. By coupling conditional generative adversarial networks (GANs) as a data augmentation technique with DECAPS, we effectively address the issue of data scarcity. Three well-trained and experienced radiologists then assessed the generated samples as a means of quality control. To further evaluate our model, we then compared its performance to the collaborating radiologists, showing that it outperformed all of them in a high-quality subset of our test set images. We then presented several activation maps to the radiologists for qualitative analysis of the classifier, and show that the information being used was relevant to the classification task. While further experimentation with a larger dataset is recommended, our results indicate that architectures like DECAPS could be used to assist radiologists in the CT mediated diagnosis of COVID-19.



\normalsize
\bibliography{references}


\end{document}